\documentclass[pre,amsmath,amssymb,letterpaper,twocolumn,10pt]{revtex4}
\pdfoutput=1

\usepackage{amsmath}
\usepackage{hyperref}
\usepackage{graphicx}
\usepackage{natbib}

\usepackage{setspace}
\usepackage{soul}

\usepackage[margin=2cm]{geometry}
\def\kt{k_{\rm B}T}

\def\cal#1{\mathcal{#1}}
\def\eqq#1{Eq.~(\ref{#1})}
\def\fig#1{Fig.~\ref{#1}}
\def\eq#1{(\ref{#1})}

\def\nm{N_{\rm max}}

\def\beq{\begin{equation}}
\def\eeq{\end{equation}}
\def\bea{\begin{eqnarray}}
\def\eea{\end{eqnarray}}

\usepackage{color}
\def\boxit#1{\vbox{\hrule\hbox{\vrule\kern6pt\vbox{\kern6pt#1\kern6pt}\kern6pt\vrule}\hrule}}

\begin{document}
\title{Patterning a surface so as to speed nucleation from solution}
\author{Lester O. Hedges}
\author{Stephen Whitelam}
\affiliation{Molecular Foundry, Lawrence Berkeley National Laboratory, Berkeley, CA 94720, USA}

\begin{abstract}
Motivated by the question of how to pattern a surface in order to best speed nucleation from solution, we build on the work of Page and Sear [Phys. Rev. Lett. 97, 65701 (2006)] and calculate rates and free energy profiles for nucleation in the 3d Ising model in the presence of cuboidal pores. Pores of well-chosen aspect ratio can dramatically speed nucleation relative to a planar surface made of the same material, while badly-chosen pores provide no such enhancement. For a given pore, the maximum nucleation rate is achieved when one of its two horizontal dimensions attains a critical length, largely irrespective of the other dimension (provided that the latter is large enough). This observation implies that patterning a surface in a raster-like fashion is a better strategy for speeding nucleation than e.g. scoring long grooves in it.
\end{abstract}

\maketitle

\section{Introduction}

The presence of a pore or a pit in a surface can affect nucleation in a profound way~\cite{Choi:2012,van2010design,Zhu:2009,Talanquer:2001,Husowitz:2004,Saugey:2005,Chayen:2006,Gelb:1999}. Page and Sear~\cite{page2006heterogeneous} used the 2d Ising model to demonstrate that nucleation of a new phase can be much faster in a pore than on a flat surface of the same material, because nucleation can start at energetically-preferred binding sites in a pore corner. Further, for given thermodynamic conditions, they showed that there exists a pore width that maximizes nucleation rate. The existence of this maximum follows immediately from the fact that as one makes a 2d pore wider, the rate for nucleation {\em into} the pore is reduced, while the rate for nucleation {\em out of} the pore (into solution) is enhanced. Here we show that similar arguments in three dimensions suggest simple strategies for patterning a surface in order to best speed nucleation from solution.

In making this claim, we present an analysis of nucleation in the Ising model that complements several previous studies. We calculate rates and free energy profiles for nucleation in the 2d Ising model in the bulk, at flat surfaces, and in the presence of rectangular pores, and in the 3d Ising model in the presence of cuboidal pores. We first calibrate our free energy sampling procedure by following Ref.~\cite{Ryu:2010} and comparing free energy profiles for bulk nucleation in the 2d Ising model with the predictions of classical nucleation theory (CNT) modified to accommodate nucleus shape fluctuations. As did Ref.~\cite{Ryu:2010}, we find excellent agreement between theory and simulation over a wide range of conditions. We next show that a flat surface can enhance nucleation, provided that this surface exerts a sufficiently large attraction (explicit or effective) for the nucleating phase. Otherwise, nucleation happens in the bulk. While intuitively reasonable, and described qualitatively by CNT, we argue that the need for such an attraction is obscured in the spin-spin representation of the Ising model, and is made clear only in the lattice gas (particle-vacancy) one. We then revisit the study of Ref.~\cite{page2006heterogeneous} by calculating free energy profiles for pore nucleation in the 2d Ising model. These profiles support and complement the nucleation rates presented in that work, showing the existence of a pore size optimum for speeding nucleation from solution. We end with our main results, rates and free energy landscapes for nucleation in the 3d Ising model in the presence of a cuboidal pore. We find that nucleation profiles display single or double barriers, depending on pore size and aspect ratio. Pores of well-chosen aspect ratio can dramatically speed nucleation relative to a planar surface made of the same material, while badly-chosen pores provide no such enhancement. For a given pore, the minimum barrier to nucleation is achieved when one of its two horizontal dimensions attains a critical length, largely irrespective of the other dimension, provided that the latter is large enough. This observation implies that patterning a surface in a raster-like fashion is a better strategy for speeding nucleation than e.g. scoring long grooves in it.

We note at the outset that there are important limitations to our study in terms of its relevance to real systems. We have chosen to work with the Ising model, a prototypical description of phase change and nucleation~\cite{stauffer1982monte,brendel2005nucleation,acharyya1998nucleation,wonczak2000confirmation,binder1974investigation,Maibaum:2008}, because it captures important effects of fluctuations and geometry crucial to many physical processes. Insights derived from it are often transferrable to many different physical systems~\cite{binney1992theory,chandler}. In particular, the result that there exists a pore size and shape in three dimensions that minimizes free energy barriers to nucleation follows immediately from geometrical considerations, in much the same way as the result that in two- and three-dimensional bulk space there exists a free energy barrier to nucleation. We expect therefore that this result should be relevant to three-dimensional systems generally. However, we do not represent important physical processes that may, in real systems, act to mask the existence of such an optimum. By representing the nucleating phase as a lattice-based, structureless one, we cannot capture potentially important effects like the mismatch in registry between a crystal and its template~\cite{van2010design}. Furthermore, our simulation protocol (grand-canonical Monte Carlo) is an efficient way of mapping free energy profiles, but ignores effects of mass transport and particle correlations that may be crucially important near real pores and surfaces. With these caveats in mind, we proceed to our study. 

\section{Model and simulation methods}

We consider homogeneous and heterogeneous nucleation in the 2- and 3-dimensional Ising model on a square or cubic lattice. Because we have in mind the nucleation of particles from solution, we find it convenient to work in the lattice gas (particle-vacancy) representation. Regardless of dimension, the energy function of our system is
\begin{equation}
\label{Eqn:lattice_gas_hamiltonian}
	E=-J \sum_{\langle ij\rangle }n_i n_j - \mu \sum_i n_i - J_{\rm{s}}\sum^{\rm wall}_{ij} n_i n_j^{\rm w}.
\end{equation}
Here $n_i=0$ if site $i$ is vacant, and $n_i=1$ if site $i$ is occupied by a particle. The first two terms are the usual bulk ones:  $J$ is the strength of the nearest-neighbor coupling, and $\mu$ is a chemical potential that can be tuned to favor particles or vacancies. The first sum runs over all distinct nearest-neighbor bulk bonds, and the second sum runs over all bulk sites. The third term describes interactions between particles and walls: this sum runs over all bonds connecting wall sites to bulk sites. Wall sites are considered to be particles, i.e. $n_j^{\rm w} =1$. Our simulations were done in the lattice gas representation, but we will use regular Ising (spin-spin) variables where convenient. For reference: via the usual mapping, $n_i = \frac{1}{2}(1+S_i)$, where $S_i = \pm1$, the bulk lattice gas maps (ignoring constant terms) to the bulk Ising model 

\begin{equation}
\label{Eqn:ising_hamiltonian}
	E_{\rm Ising}^{\rm bulk}=-K \sum_{\langle ij\rangle }S_i S_j - h\sum_i S_i,
\end{equation}
where $J = 4K$ and $\mu = 2h-2z K$. Here $z=2d$ is the coordination number of the $d$-dimensional square or cubic lattice.

We carried out simulations using a standard grand canonical Metropolis Monte Carlo (MC) procedure~\cite{frenkel1996understanding}. Each trial move consisted of an attempted change of  state of a randomly-chosen lattice site. Trial moves resulting in an energy change $\Delta E$ were accepted with probability $\mathrm{min}\left(1,\mathrm{exp}(-\beta \Delta E)\right)$, where $\beta \equiv 1/(k_{\rm{B}}T)$. To study homogeneous nucleation in 2d, we used a system of $N_{\rm{box}}=100^2$ sites with periodic boundary conditions in both directions. To model a wall we use periodic boundaries in the horizontal direction only, and created a closed boundary along the top and bottom of the box by filling all sites in the first row of the lattice with particles. Following the protocol of Ref.~\cite{page2006heterogeneous}, we also simulated rectangular pores in a system of $60^2$ sites. The depth of a pore  was fixed at 30 sites and its width was allowed to vary. An analogous protocol was used in 3d simulations to generate cuboidal pores of fixed depth and a range of different length-to-width aspect ratios.

To calculate free energy landscapes for nucleation we used standard umbrella sampling~\cite{torrie1977nonphysical} protocols, and for the calculation of nucleation rates we used forward flux sampling (FFS) \cite{Allen:2006}. Brief details of both methods are given in Appendix~\ref{apx:umbrella_sampling}.

\section{Results: 2d simulations}
\subsection{Free energy barriers for bulk nucleation in the 2d Ising model are well described by (modified) classical nucleation theory}
\label{Section:ising_bulk}
\begin{figure*}[!htb]
\center
\includegraphics[width=0.95 \linewidth]{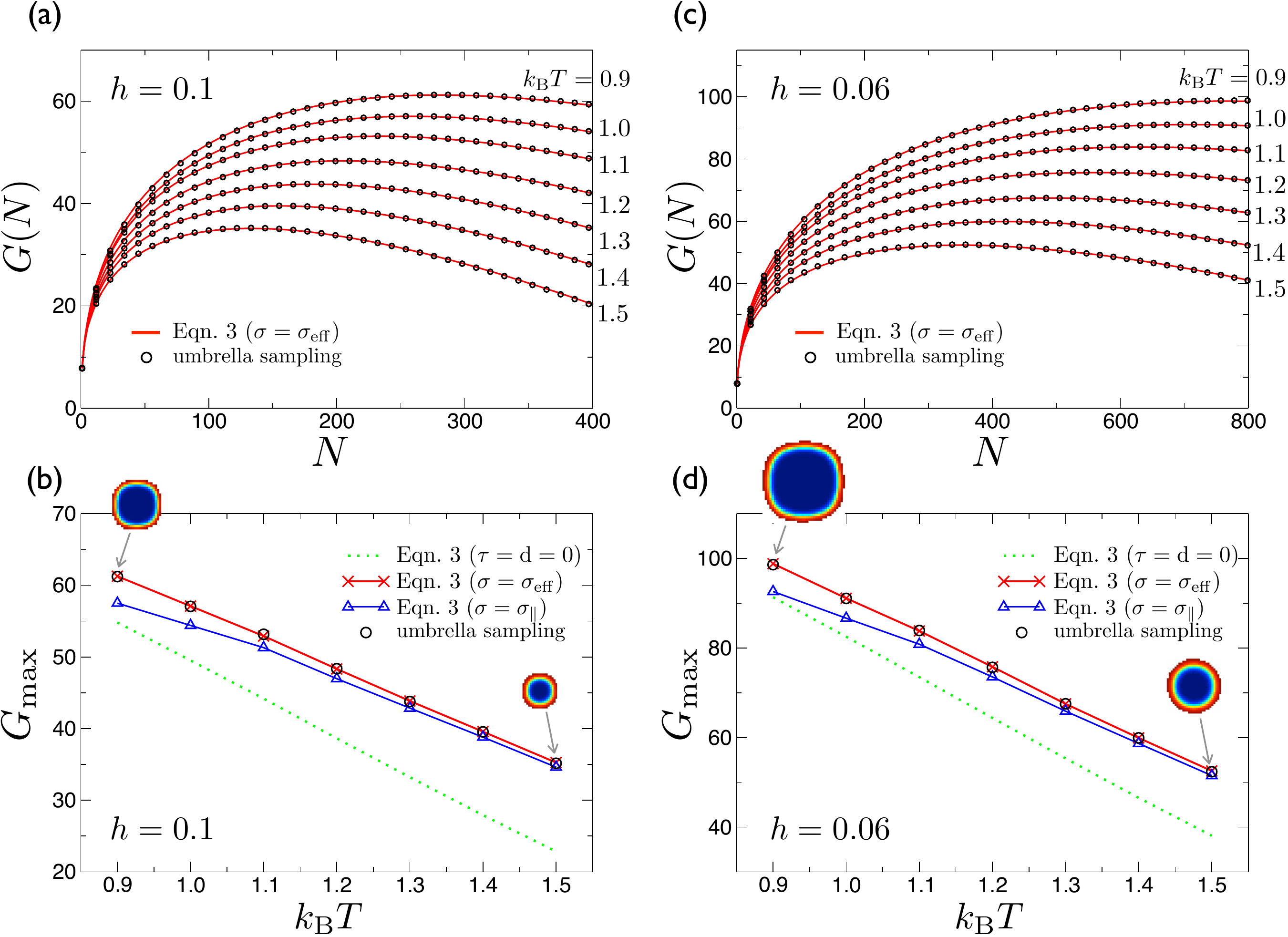}
\caption{Free energy barriers for homogeneous nucleation in the 2d Ising model are well-described by~\eqq{Eqn:ising_free_energy_corrected}. We start from an empty box (a box of vacancies), and use umbrella sampling to grow clusters of connected particles (droplets). Panels (a) and (c) show droplet free energy profiles as function of droplet size $N$, for various values of $\kt$ and for magnetic field strengths $h=0.1$ and $h=0.06$. Open circles denote the results of umbrella sampling simulations; solid lines are obtained from~\eqq{Eqn:ising_free_energy_corrected}, using the Shneidman et al. effective surface tension. In agreement with Ref.~\cite{Ryu:2010}, the theoretical predictions match the simulation data closely. Panels (b) and (d) compare free energy barriers obtained from simulation in the plots directly above them. The dotted lines show the conventional (uncorrected) CNT barrier height predictions of~\eqq{Eqn:ising_free_energy_corrected} with $\tau=d=0$. In all cases the uncorrected theory significantly underestimates the barrier height. Upward pointing triangles denote the results of~\eqq{Eqn:ising_free_energy_corrected} using the Onsager solution to the surface tension. Agreement is good at high temperature but less good at low temperature, where droplets are anisotropic (see snapshots on diagram).}
\label{Fig:ising_bulk}
\end{figure*}

By way of calibration, it is instructive to compare nucleation free energy barriers calculated by umbrella sampling with the predictions of classical nucleation theory~\cite{becker1935kinetische,volmer1926keimbildung} (CNT). Ref.~\cite{Ryu:2010} showed that sampled free energy barriers in the 2d Ising model are in excellent agreement with the predictions of the CNT-like expression
\begin{equation}
\label{Eqn:ising_free_energy_corrected}
	G_{\rm CNT}(N)=- \Delta g N + 2\sigma \sqrt{\pi N}+\tau\kt  \ln N + d(T).
\end{equation}
Here $G$ is the excess free energy of a droplet of $N$ `up' spins in a background of `down' spins. The first term of this expression is the conventional bulk reward for growth of a circular droplet. $\Delta g$ is the bulk free energy difference between the two bulk phases, equal to $2h$ at low temperature. At higher temperature (still below the critical one) the viable bulk phases are less dense than the all-up and all-down spin limits. Here we expect $\Delta g \approx h \Delta m$ to be a reasonable approximation, where $\Delta m$ is the magnetization difference between bulk phases. For the conditions considered in this section, these estimates differed by at most about $3 \%$; we therefore set $\Delta g = 2h$. The second term of~\eqq{Eqn:ising_free_energy_corrected} is the surface tension penalty for growth of a circular droplet. $\sigma$ is the inter-phase surface tension. In 2d, the Ising model surface tension (at $h=0$) in the direction of either lattice vector is known from the Onsager solution~\cite{Onsager:Ising}, and is
\begin{equation}
\label{Eqn:onsager_sigma}
	\sigma_{\parallel} = 2K -\kt \ln \coth(\beta K).
\end{equation}
Because a non-square droplet cannot be accommodated perfectly on a square lattice, it is also useful to consider the orientationally-averaged effective droplet surface tension $\sigma_{\rm{eff}}(T)$ derived by Shneidman et al.~\cite{Shneidman:1999}:
\begin{equation}
\label{Eqn:shneidman_sigma}
\sigma_{\rm{eff}}(T)\simeq \frac{1}{2\sqrt{\chi(T)}}\left(\sigma_{\parallel} + \sigma_{\rm{diag}}\right),\ \ \ T\gtrsim0.25T_{\rm{c}}.
\end{equation}
Here
\begin{equation}
	\chi(T) = \left(1 - \sinh^{-4}(2\beta K)\right)^{1/8},
\end{equation}
and
\begin{equation}
\sigma_{\rm{diag}} = {\sqrt 2} \kt \, \ln \sinh(2\beta K)
\end{equation}
is the surface tension in the direction of the unit cell diagonal \cite{Fisher:1967}. In what follows we compare our simulations with the predictions of~\eqq{Eqn:ising_free_energy_corrected} using $\sigma$ defined both by Eqns.~\eq{Eqn:onsager_sigma} and \eq{Eqn:shneidman_sigma}. 

The third term of~\eqq{Eqn:ising_free_energy_corrected} accounts for shape fluctuations of the droplet, and can be derived from field theoretic considerations of nucleation rates~\cite{Langer:1967,Lowe:1980,Gunther:1980,Gunther:1994}. The shape fluctuation parameter $\tau=5/4$ in 2d~\cite{Jacucci:1983}. One important contribution of Ref.~\cite{Ryu:2010} was to recognize that this term can be considered a contribution to the free energy of a droplet. Without such a contribution, CNT and umbrella sampling are in quantitative disagreement. The final term of~\eqq{Eqn:ising_free_energy_corrected} accounts for the fact that the conventional CNT expression has no clear origin of free energy, because it does not resolve the monomer constituents of droplets. Instead, one can fix the origin of free energy profiles in the 2d Ising model by requiring that~\eqq{Eqn:ising_free_energy_corrected} returns the free energy of (say) clusters of size 1. The latter quantity can be calculated exactly in the Ising model, so fixing $d(T)$~\cite{Ryu:2010}.

In \fig{Fig:ising_bulk} we compare free energy profiles computed from umbrella sampling simulations with~\eqq{Eqn:ising_free_energy_corrected}. Panels (a) and (c) show droplet free energy profiles for various values of $\kt$, for magnetic field strengths $h=0.1$ and $h=0.06$. Open circles denote the result of umbrella sampling simulations; solid lines are obtained using \eqq{Eqn:ising_free_energy_corrected} with the Shneidman et al. effective surface tension. As per Ref.~\cite{Ryu:2010}, the theoretical prediction fits the simulation data well across the range of conditions studied. Panels (b) and (d) compare free energy barriers obtained from simulation and theory in the plots directly above them. Here the dotted lines indicate the barrier height predictions of the uncorrected CNT expression,~\eqq{Eqn:ising_free_energy_corrected} with $d=\tau=0$. In all cases the uncorrected theoretical prediction underestimates the  barrier height. Upward pointing triangles show the barrier height prediction of~\eqq{Eqn:ising_free_energy_corrected} using the Onsager solution for the surface tension. Agreement between it and simulation is good at high temperature, but not at low temperature, where droplets are anisotropic.  Also shown are averaged droplet profiles for critical nuclei at the extreme values of $\kt$ at each value of $h$. These were obtained by averaging over all configurations for which $N=N_{\rm c}$ during umbrella sampling. Snapshots are scaled relative to the size of the largest cluster shown (which occurs when $\kt=0.9$ and $h=0.06$). Droplets are noticeably non-circular at low temperatures. 
 
\begin{figure}[]
\center
\includegraphics[width=\columnwidth]{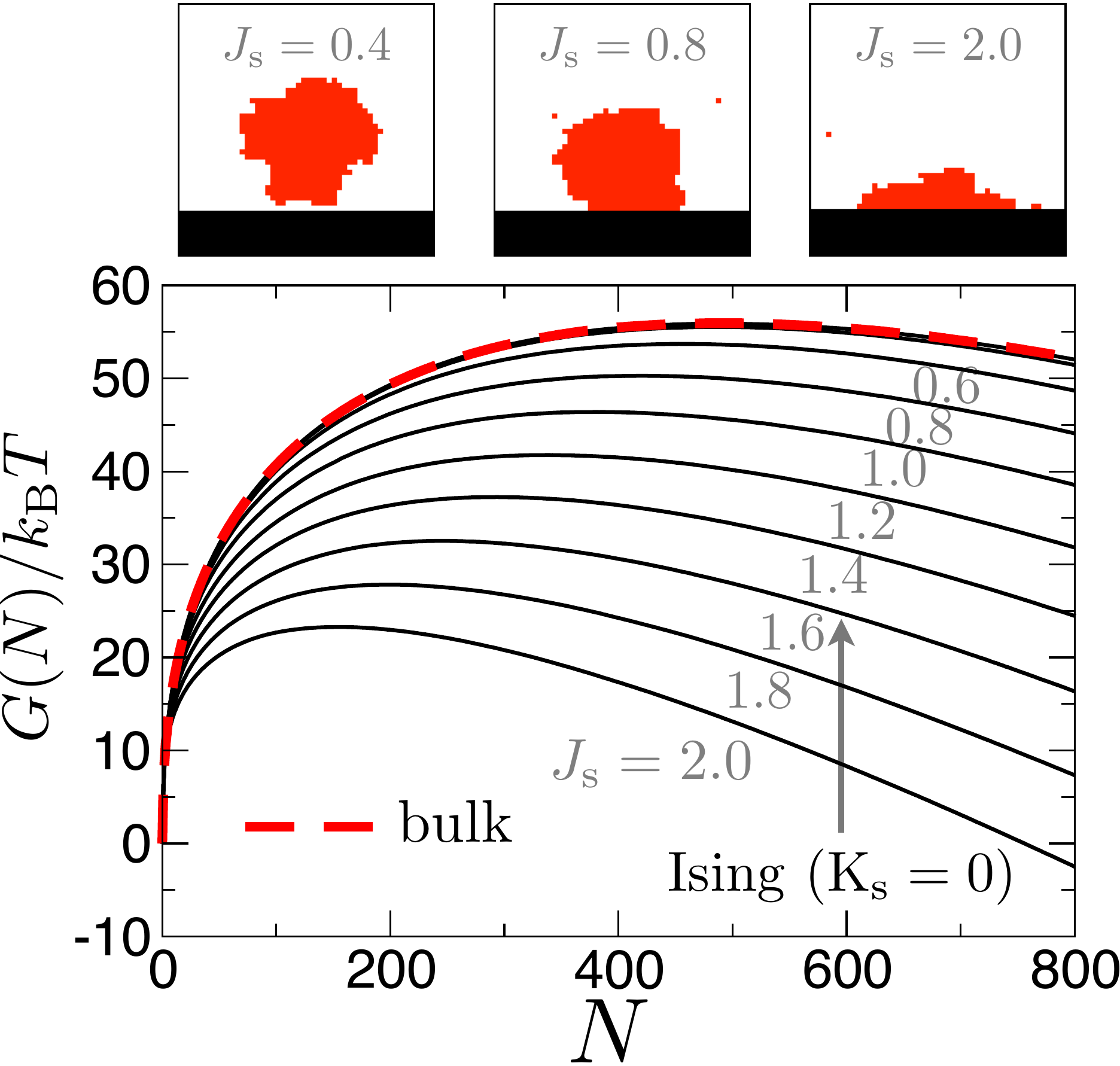}
\caption{Free energy profiles for growing a droplet in the presence of a wall, in the 2d Ising model (lattice gas representation), for various particle-wall attraction strengths $J_{\rm s}$. Nucleation in the absence of a particle-wall attraction is as in the bulk: curves for $J_{\rm{s}}=0$, 0.2 and 0.4 coincide with the bulk curve (dashed red line). The nucleation barrier and critical nucleus are reduced for large enough particle-wall attraction. The arrow marks the lattice gas particle-wall attraction $(J_{\rm s}= J/2)$ that is equivalent to the Ising model representation in the presence of an inert wall ($K_{\rm s} = 0$; see main text for discussion). Snapshots above are representative of critical nuclei at the particle-wall interactions specified.}
\label{Fig:ising_surface}
\end{figure}

\subsection{Nucleation at a planar surface is faster than in the bulk only if the surface is sufficiently attractive}

With confidence in our sampling protocol (Appendix~\ref{apx:umbrella_sampling}) established, we next turn to the question of how a planar surface affects nucleation. \fig{Fig:ising_surface} shows free energy barriers to nucleation for the 2d lattice gas (\eqq{Eqn:lattice_gas_hamiltonian}) in the presence of a flat wall, for a range of particle-wall interaction strengths $J_{\rm{s}}$. We set $J=3.2,\ \mu=-6.3$ (equivalent to $K=0.8,\ h=0.05$ in the Ising model representation) and $\kt=1$. Under these conditions the particle (up spin) phase is thermodynamically preferred to the initial vacancy (down spin) phase.  In the absence of a particle-wall attraction, particles are effectively repelled by the wall, for reasons of entropy (sites available in the bulk exceed those available near the wall) and geometry (the nucleus shape that minimizes the surface-to-area ratio in two dimensions, a circle, can form only in the bulk). For particle-wall attractions not strong enough to overcome the entropic penalty of wall confinement, nuclei again grow in the bulk of the simulation box. (Sampling was initialized using wall-hugging droplets generated using large $J_{\rm s}$; when bulk nucleation was preferred, droplets moved away from the wall). For sufficiently large attractions nuclei {\em do} grow at the wall, and the free energy barrier to nucleation and the size of the critical nucleus are smaller than their bulk counterparts. A substantial particle-wall attraction is needed to counter the favorable entropy associated with bulk nucleation: in other words, a surface will enhance nucleation only if it possesses a sufficiently strong attraction for the nucleating phase.

While this observation is intuitively reasonable, we note that it is much more apparent in the lattice gas representation than the Ising one. The authors of Refs.~\cite{page2006heterogeneous,curcio2010energetics} studied nucleation using the Ising Hamiltonian \eqq{Eqn:ising_hamiltonian} augmented by a bulk-wall interaction $E_{\rm Ising}^{\rm wall}= - K_{\rm s}\sum^{\rm wall}_{ij} S_i S_j^{\rm w}$, for the particular case $K_{\rm s}=0$ (i.e. an energetically inert wall). Although $K_{\rm{s}}=0$ means that the wall has no energetic preference for either phase, the {\em repulsion} between unlike spins in the bulk leads to an effective attraction between the nucleating phase (in those papers the up-spin phase) and the wall~\footnote{An inert wall in the Ising model representation preserves the up-down symmetry of the system, favoring neither phase. However, it enhances nucleation of the new thermodynamic phase. Flipping an up spin at a planar interface between up- and down spins costs $4K + 2h$ in energy. By contrast, if a bulk up phase lies in contact with an inert wall, flipping down a spin in contact with the wall costs a greater amount, $6K + 2h$: the wall stabilizes, energetically, the phase that wets it. In the lattice gas representation, by contrast, an energetically inert wall favors the vacancy phase over the particle phase. A particle-wall attraction of strength one-half of the particle-particle bulk coupling is required to restore the symmetry intrinsic to the Ising model.}. Carrying through the Ising-lattice gas transformation, it can be shown that an Ising model in contact with an energetically inert wall $(K_{\rm s}=0)$ is equivalent to a lattice gas in contact with a wall that possesses a substantial interaction for particles, i.e. \eqq{Eqn:lattice_gas_hamiltonian} with $J_{\rm s}=J/2$ \cite{Sear:Email2011}. This limit is marked by an arrow in \fig{Fig:ising_surface}. In both cases, one must engineer a substantial attraction between the nucleating phase and a wall before nucleation happens at the wall in preference to in bulk (a result confirmed by simple scaling arguments: see Appendix~\ref{sec:plane}). In the lattice gas representation the coupling $J_{\rm s}$ might be regarded as a literal wall-particle attraction; in Ising language, the attraction between up spins and the wall can be regarded as an effective one, mediated by `solvent' (the down-spin phase).

\subsection{Nucleation in 2d pores can be faster still}
\label{sec:2d_pores}

\begin{figure*}[!htb]
\center
\includegraphics[width=\linewidth]{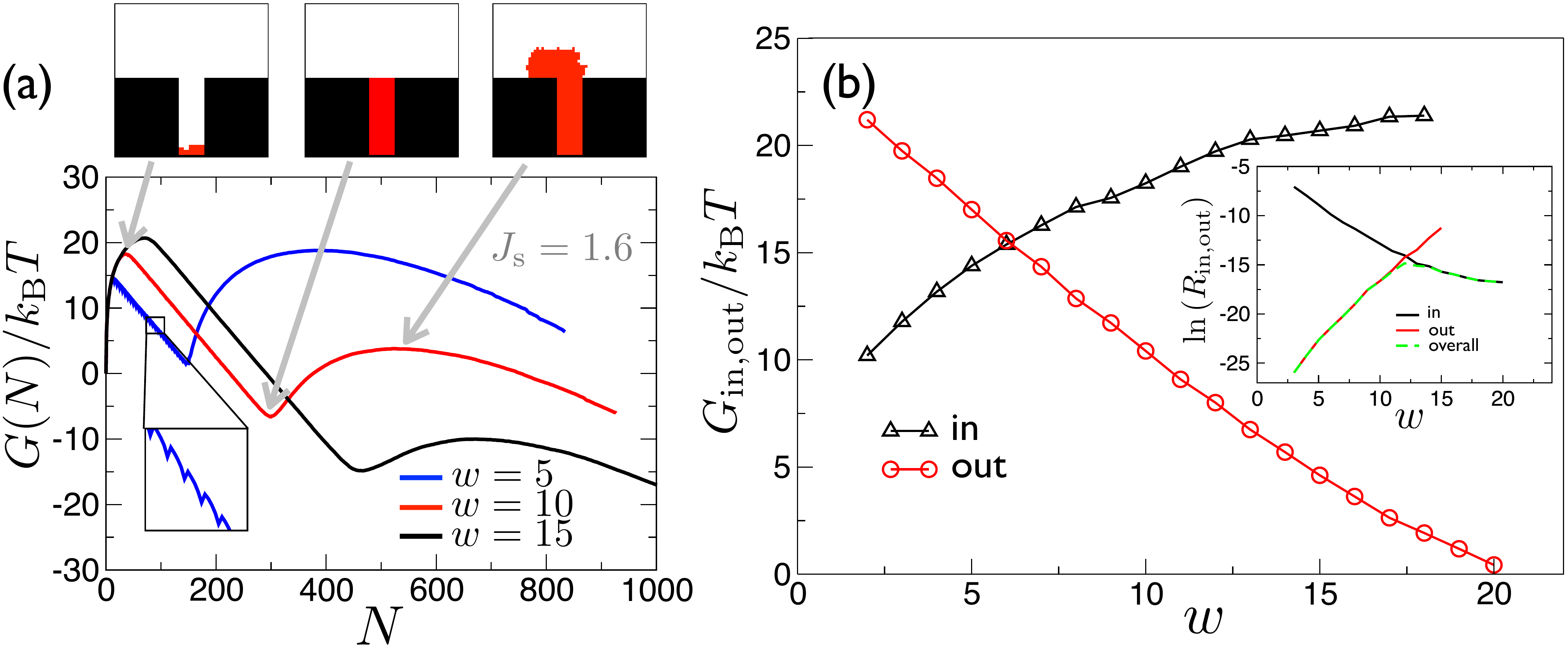}
\caption{(a) Free energy profiles for nucleation in 2d Ising model pores. Profiles are consistent with the two-step mechanism described in Ref.~\cite{page2006heterogeneous}, revealing a barrier to nucleation into the pore, and a barrier to nucleation from a filled pore into the bulk (see snapshots). The width $w$ of the pore governs the heights of the two barriers, which show opposing dependencies on $w$. The total nucleation rate is a competition between these two processes and is optimized, for given thermodynamic conditions, by a specific pore width. The expanded region of the blue ($w=5$) free energy curve illustrates that completed rows represent local metastable minima during the post-critical filling of a pore. (b) Barrier to nucleation inside a pore (black triangles) and out of filled pore (red circles) as a function of the width of the pore $w$. Note that the pore width that maximizes nucleation {\em rate} cannot be determined directly from the intersection of the two curves; instead, one must compute nucleation rates explicitly. The inset shows nucleation rates, $R_{\rm{in,out}}$, computed using the forward flux sampling method \cite{Allen:2006}. In agreement with the results of Ref.~\cite{page2006heterogeneous}, we find that the overall nucleation rate (the reciprocal of the sum of the `in' and `out' nucleation timescales) is optimized for a pore about 12 sites wide.}
\label{Fig:ising_pore}
\end{figure*}

\begin{figure}[!b]
\center
\includegraphics[width=0.9 \columnwidth]{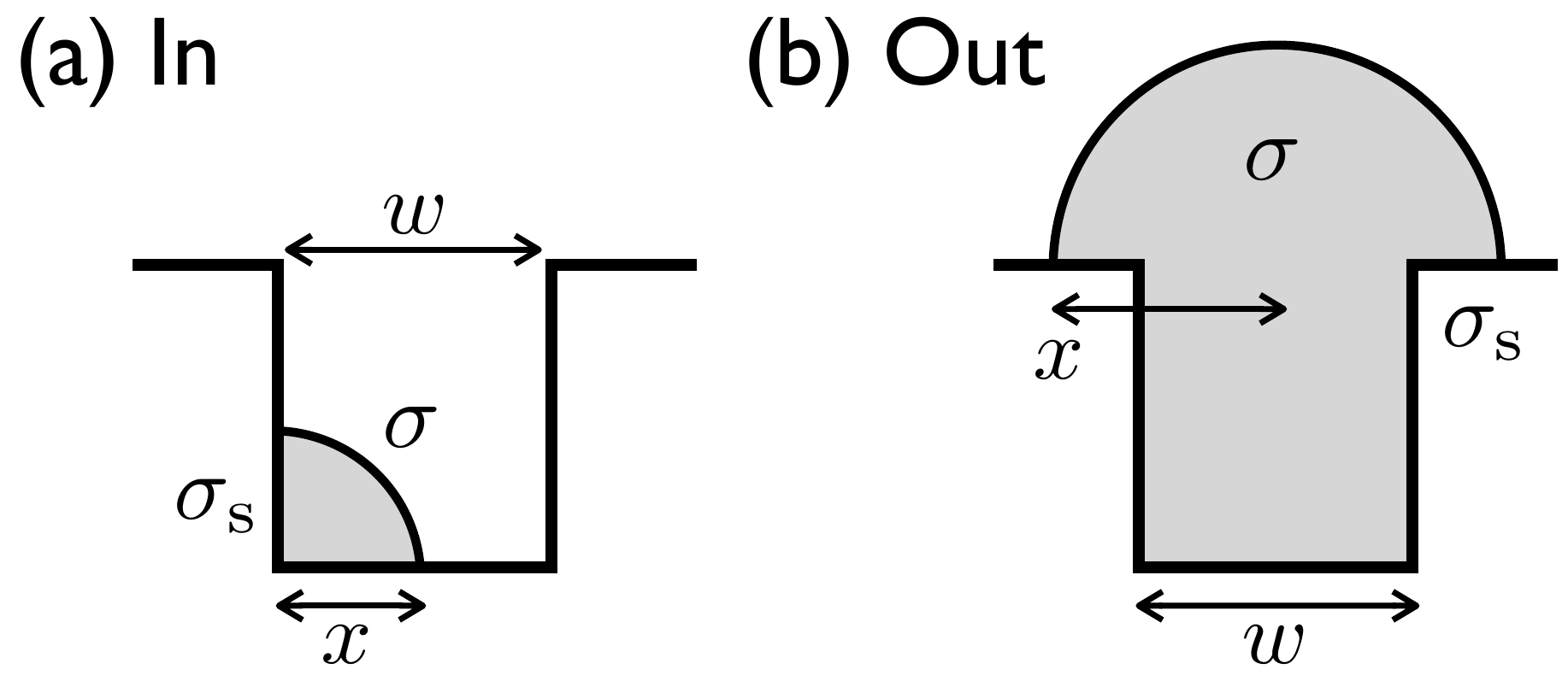}
\caption{Geometry for CNT-like scaling argument for pore nucleation~\cite{page2006heterogeneous}: (a) quarter-circular droplet nucleating in a pore corner, and (b) semicircular droplet nucleating out of a filled pore. Simple approximations for the free energy barriers in cases (a) and (b) show that the barrier for nucleation within a pore increases with $w$, while the barrier for nucleation out of a pore decreases with $w$. This competition implies the existence of an optimal pore with.}
\label{Fig:cnt_geometry_pore}
\end{figure}

If a surface is sufficiently attractive, then, it can render nucleation faster than in the bulk. Nucleation in a pore made out of that surface can be faster still, because pore corners provide a convenient initiation site for the new thermodynamic phase~\cite{page2006heterogeneous}.  Further, for given thermodynamic conditions, Ref.~\cite{page2006heterogeneous} demonstrated that there exists a pore size that maximizes nucleation rate. The existence of this maximum follows from the fact that as one makes a pore bigger, the rate for nucleation {\em into} the pore is reduced, while the rate for nucleation {\em out of} the pore (into solution) is enhanced. In~\fig{Fig:ising_pore}(a) we show free energy profiles for nucleation in a 2d pore. These profiles complement the nucleation rate calculations of Ref.~\cite{page2006heterogeneous}, confirming the existence of a double barrier to pore-mediated nucleation into solution. Under our umbrella sampling protocol, nucleation first occurs within the pore (starting in one of the corners due to the greater number of favorable energetic contacts there), and is followed by nucleation from a filled pore into the bulk. As highlighted in the boxed region, free energy profiles for narrow pores show local metastable minima during the post-critical filling of a pore. Each minimum corresponds to a filled row of the pore.
 \begin{figure*}[t!]
\center
\includegraphics[width=\linewidth]{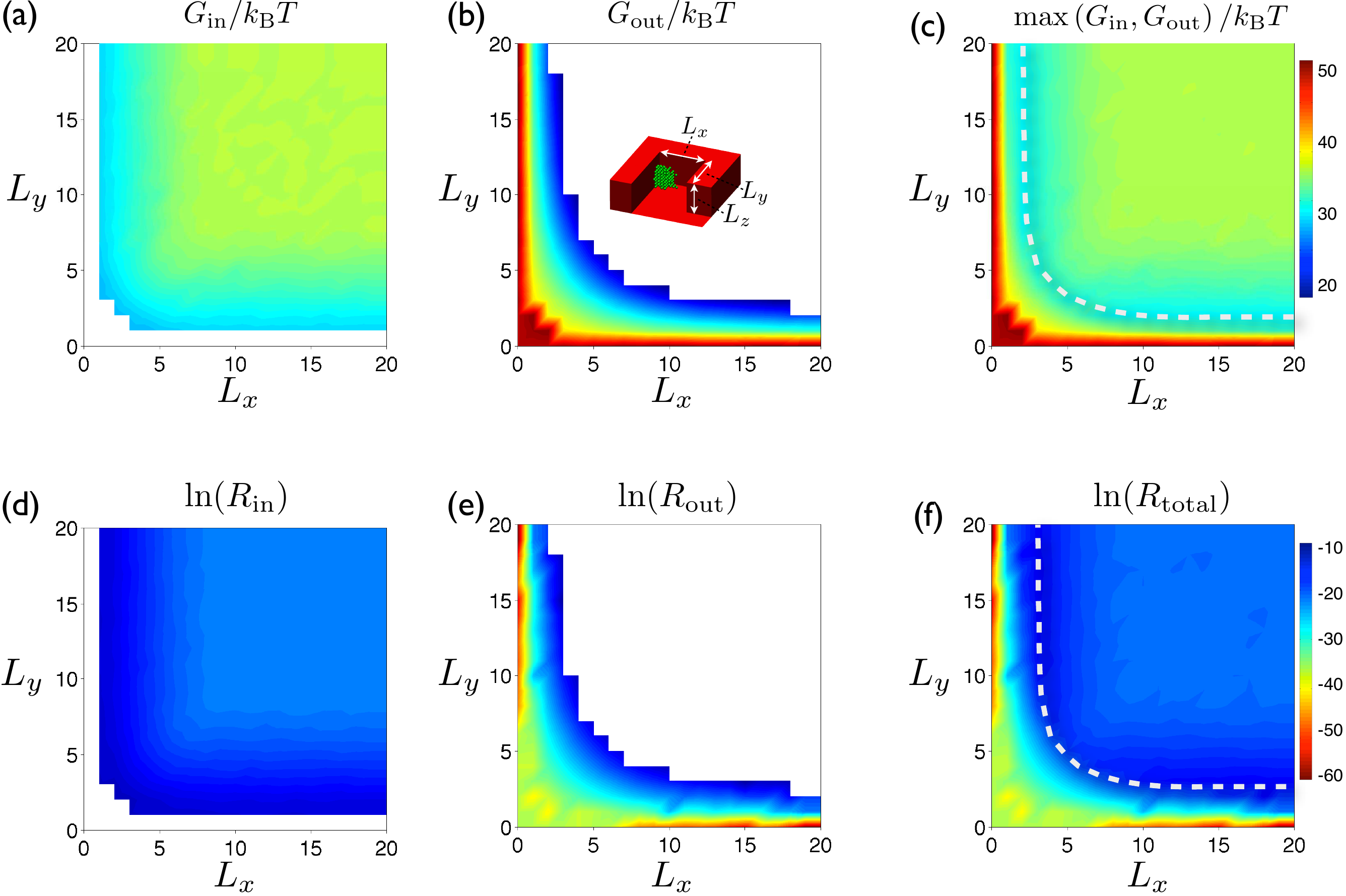}
\caption{Heterogeneous nucleation in three dimensions in the presence of a cuboidal pore of $L_z = 10$ lattice sites and aspect ratio $L_x:L_y$. (a) Contour map showing the barrier to nucleation inside a pore. Within the white region there is no stable nucleus within the pore. (b) Contour map showing barrier to nucleation out of a pore. Within the white region there is no barrier to nucleation out of the pore, i.e. a critical nucleus can form within the pore and grow without bound. (c) Contour map of the largest barrier to nucleation (either in pore, out of pore, or single barrier). Since nucleation occurs in a pore corner, and the number of corners does not change with increasing pore size, the nucleation barrier is strongly dependent upon only a single horizontal dimension. If   one pore dimension takes this `correct' size, the nucleation barrier depends only weakly upon the other pore dimension, provided that the latter is large enough. The region of optimum pore geometry therefore forms a well-defined band, as highlighted by the dashed white line. The color scale is the same for all plots. Panels (d)--(f) show complementary contour maps of the nucleation rate (calculated using forward flux sampling) for the panel directly above. The low lying band seen in the panel (c) corresponds, to within a lattice site, to a ridge in panel (f), along which the overall rate of nucleation is maximized.}
\label{Fig:hot_spot}
\end{figure*}

As described in Ref.~\cite{page2006heterogeneous}, the barrier to nucleation within the pore increases with increasing pore width, while the barrier to nucleation out into solution shows the opposite trend (see panel (b)). It should be noted that the pore width that maximizes nucleation {\em rate} cannot be determined directly from the intersection of the two curves: one must compute droplet nucleation rates explicitly. The inset shows the two nucleation rates, $R_{\rm{in,out}}$, in units of Monte Carlo steps per bulk lattice site, computed using forward flux sampling~\cite{Allen:2006}. In agreement with the results of Ref.~\cite{page2006heterogeneous} we find that the overall nucleation rate (the reciprocal of the sum of the `in' and `out' pore nucleation timescales) is largest for a pore approximately 12 sites wide.

A simple CNT-like approximation confirms that, given a sufficiently strong particle-wall attraction, there must exist a pore size that maximizes nucleation rate (this argument is that of Ref.~\cite{page2006heterogeneous}, modified to account for variable particle-wall surface tension). Approximating the in-pore nucleus as a quarter-circle droplet of radius $x$ growing from the corner of a pore of width $w$ (see \fig{Fig:cnt_geometry_pore}(a)) suggests a free energy cost for the nucleus of $G_{\rm{in}}(x)\simeq 2x\sigma_{\rm{s}} + {\pi\sigma x}/2 -\Delta g x^2 \pi/4$ (here $\sigma_{\rm s}$ is the droplet-wall surface tension). Assuming that the pore is narrow (so that the function $G_{\rm{in}}(x)$ does not reach its turning point for $x < w$) then the barrier to nucleation goes as $G_{\rm{in}}^{\rm{max}}(w) \sim (2\sigma_{\rm{s}} + \pi\sigma/2)w-{\cal O}(w^2)$, which {\em increases} (sub-linearly) with pore width $w$. For nucleation out of a filled pore (\fig{Fig:cnt_geometry_pore}(b)), by a semi-circular droplet radius $x$, the free energy profile can be approximated as $G_{\rm{out}}(x)\simeq (2x-w) \sigma_{\rm s} +\pi x \sigma -\Delta g x^2 \pi/2$, which gives a pore width-dependent nucleation barrier of $G_{\rm{out}}^{\rm{max}}(w)=(2\sigma_{\rm{s}} + \pi\sigma)^2/(\pi\Delta g) - w\sigma_{\rm_{s}}$. This {\em decreases} linearly with $w$ (note that the `out' barriers seen in Fig.~\ref{Fig:ising_pore} are indeed approximately linear in $w$). An optimum pore width arises naturally from the competition between these two processes.

We stress, however, that the degree of attenuation of the nucleation barrier due to the pore depends on the particle-pore attraction, and can range from nothing at all (for small $J_{\rm s}$), to total (for large $J_{\rm s}$). 

\section{Results: 3d simulations}
\subsection{Nucleation barriers in 3d pores depend on pore size and aspect ratio}

\begin{figure*}[]
\center
\includegraphics[width=\linewidth]{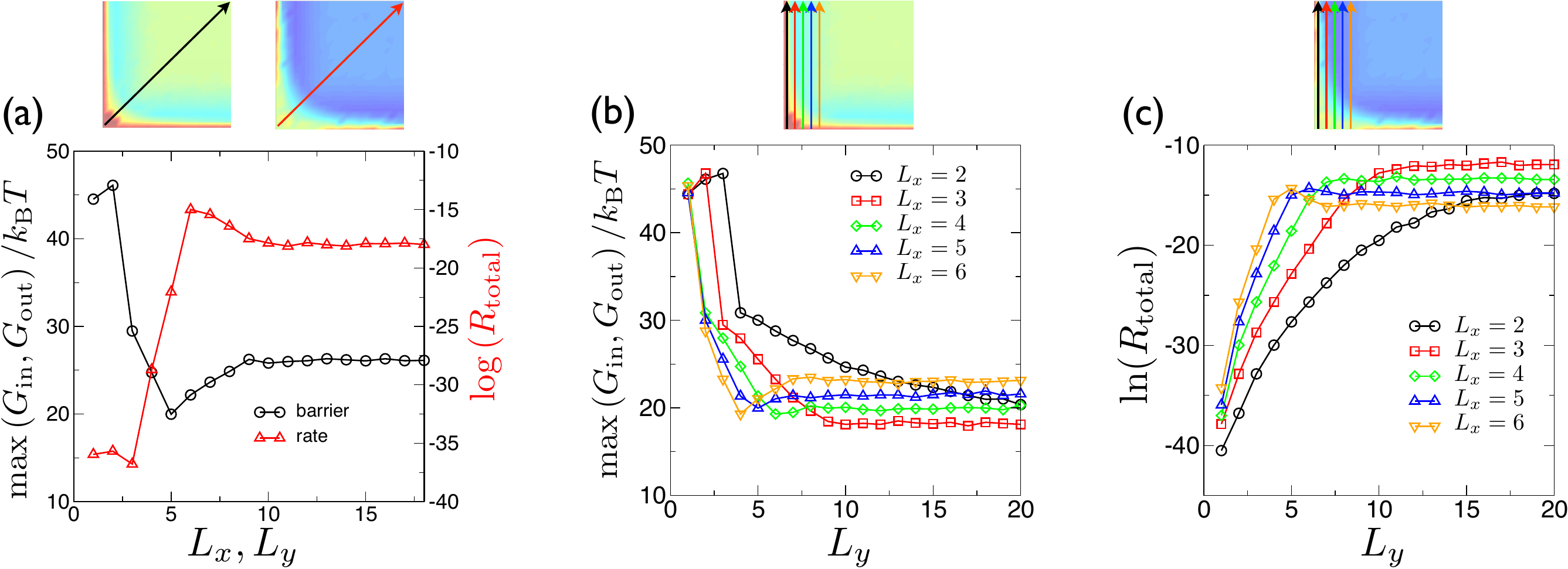}
\caption{(a) Largest barrier height and total nucleation rate taken from diagonal cuts ($L_x=L_y$) along contour maps (c) and (f) in \fig{Fig:hot_spot}. The square aspect ratio pore that maximizes the nucleation rate has a side of length $L_x=L_y=6$. (b) If $L_x$ takes a certain, optimal value, then the barrier height depends only weakly on $L_y$, provided that $L_y$ is large enough. (c) The same is true of the overall nucleation rate.}
\label{Fig:hot_spot_cuts}
\end{figure*}

\begin{figure*}[!htb]
\center
\includegraphics[width=0.95\linewidth]{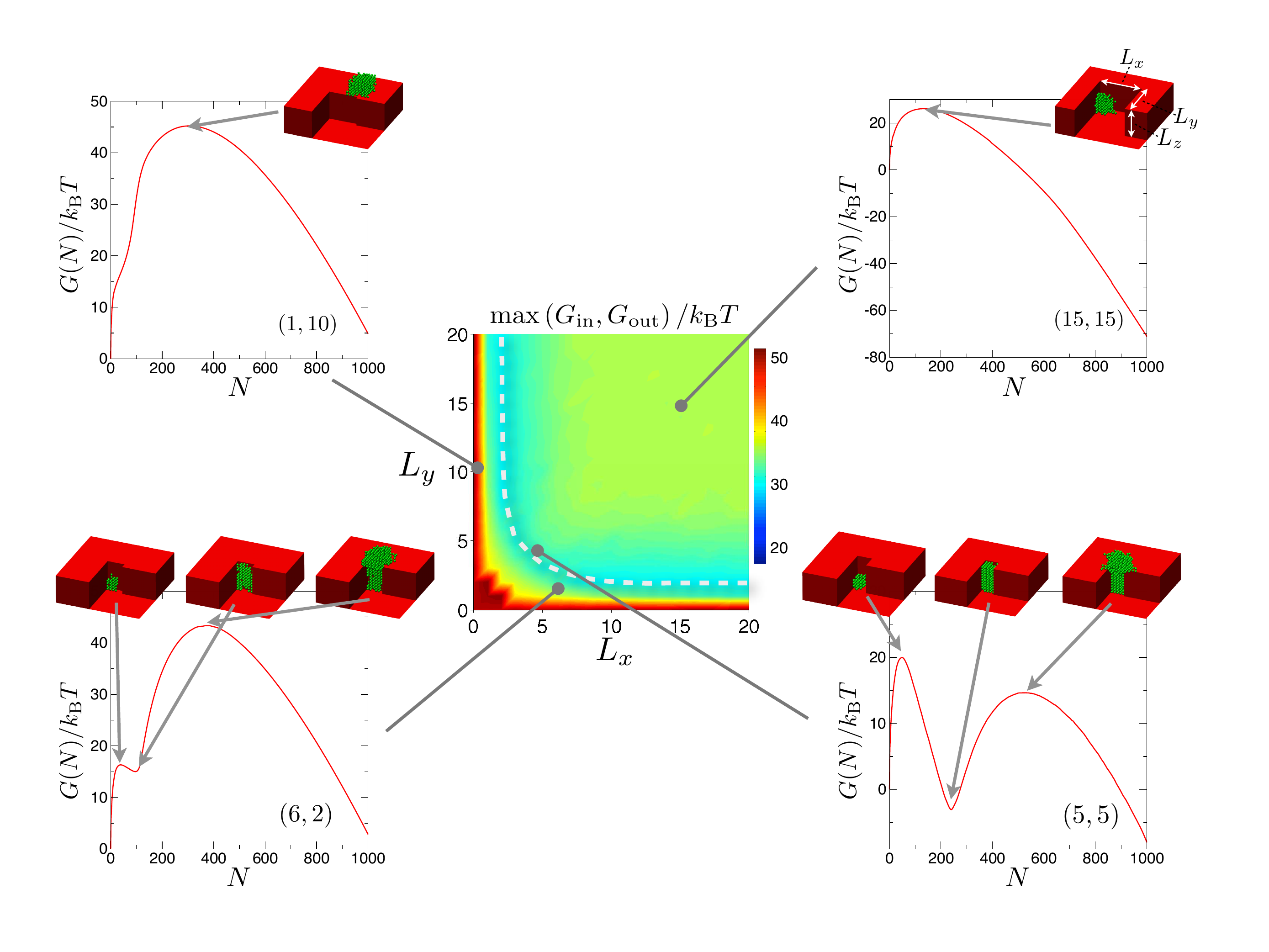}
\caption{Nucleation mechanism as a function of pore aspect ratio $(L_x,L_y)$ for pores of fixed depth $L_z=10$. Snapshots and free energy profiles illustrate the key behaviors seen: (top left) droplets are unstable within the pore,  and so a critical nucleus can only form on the flat surface instead; (top right) the pore is so large that a droplet within it becomes critical before the pore is filled; (bottom left) the filled pore is metastable with respect to the empty lattice (solution) and full lattice (stable, nucleated phase); (bottom right) the filled pore is stable with respect to solution and metastable with respect to the nucleated phase. For clarity, snapshots show only particles within the largest cluster in the simulation box.}
\label{Fig:hot_spot_breakdown}
\end{figure*}

The analysis of Ref.~\cite{page2006heterogeneous} can be straightforwardly extended to treat cuboidal pores. We studied heterogeneous nucleation in a three-dimensional lattice gas of $30^3$ sites, for  $J=1.6,\ \mu=-3.5$, in the presence of a cuboidal pore of dimensions $L_x \times L_y \times L_z$ embedded in a surface. Both the pore and the surface possessed stickiness parameter $J_{\rm s} = 0.6$. For a fixed pore depth of $L_z=10$ sites we computed nucleation free energy profiles. \fig{Fig:hot_spot} demonstrates that, in general, there exists a barrier to filling a pore ($G_{\rm in}$, \fig{Fig:hot_spot}(a)), and for growing a droplet from a filled pore into solution ($G_{\rm out}$, \fig{Fig:hot_spot}(b)). For very large pores, however, droplets can attain criticality in the pore, and no barrier to growing into solution exists. Very small pores, by contrast, cannot be filled without substantial free energy cost, and offer no enhancement of nucleation relative to a planar surface. When such pores are present, nuclei grow on the surrounding flat substrate (made of the same material as the pore).
\begin{figure*}[ht]
\center
\includegraphics[width=\linewidth]{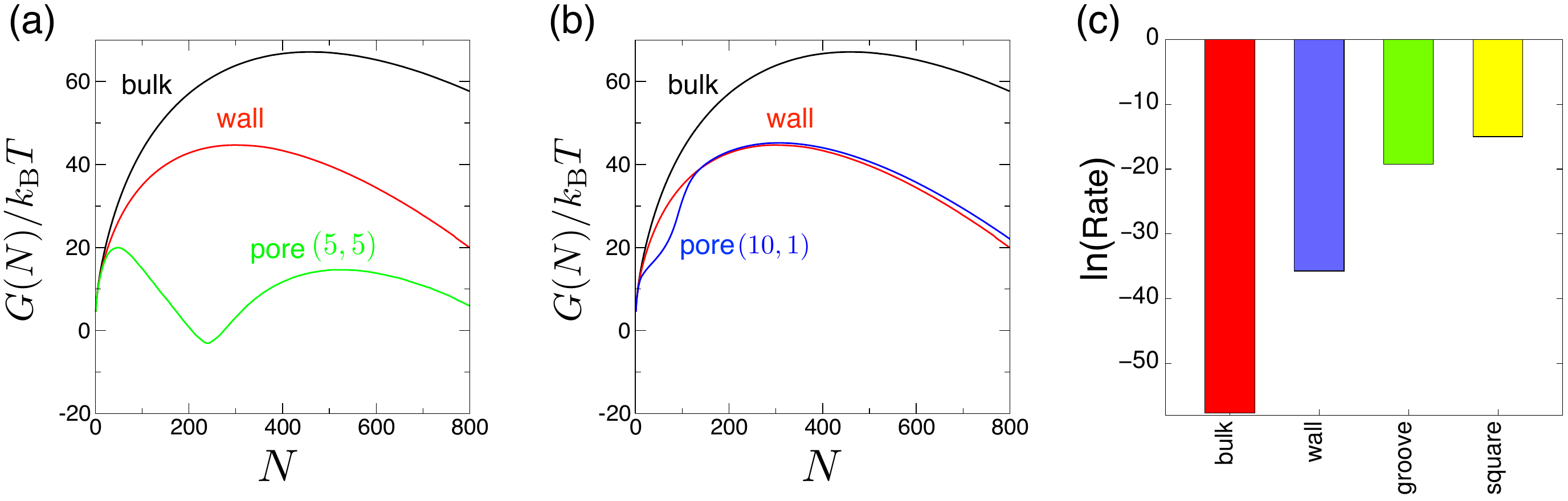}
\caption{Well-chosen pores can dramatically quicken nucleation relative to a planar surface made of the same material, while badly-chosen pores do little except take up space on the substrate. (a) If a planar surface has a sufficient attraction for the nucleating phase then it can promote nucleation relative to the bulk case. Here the bulk nucleation barrier is  about $ 67\, \kt$; at an attractive wall ($J_{\rm s}=0.6$) the barrier is instead $45\, \kt$. A pore of well-chosen geometry can reduce the barrier even further. Here nucleation within a square $5 \times 5$ pore has a maximum barrier of about $20\, \kt$. (b) By contrast, a badly-chosen pore offers no enhancement of nucleation over a flat surface of the same material. Given a pore so narrow that droplets are unstable within it, nucleation takes place instead on the surrounding surface. (c) Summary: nucleation at a planar surface (of stickiness $J_{\rm s}=0.6$)  is 9 orders of magnitude faster than in the bulk. A long groove of optimum width promotes nucleation rate by a further 7 orders of magnitude. A square pore of optimum width gives rise to nucleation that is about 70 times faster still, suggesting that, for given thermodynamic conditions, a raster arrangement of pores -- repeating copies of the square -- is a better way to speed nucleation than e.g. scoring long grooves in the surface.}
\label{Fig:behavior_range}
\end{figure*}

\fig{Fig:hot_spot}(c) shows a contour plot of the larger of the `in' and `out' barriers, $G_{\rm max}$, as a function of a pore's horizontal dimensions. All pores provide a significant attenuation of the free energy barrier to nucleation relative to bulk, because the pore and its surrounding surface is attractive (here the bulk nucleation barrier is $\approx 67\, \kt$). However, $G_{\rm{max}}$ varies dramatically with pore geometry. Achieving maximum attenuation of the nucleation barrier requires making only one of the pore's horizontal dimensions the `correct' size, provided that the second is large enough. Consequently, the region of optimal pore geometry forms a well-defined band on the contour plot. Panels (d)-(f) show complementary contour plots of nucleation rates, computed using forward flux sampling. The band seen in panel (c) corresponds, to within a lattice site, to a ridge in panel (f), along which the overall rate of nucleation is maximized. 

A selection of cuts along the contour plots (corresponding to varying either the size or shape of a pore) are shown in \fig{Fig:hot_spot_cuts}. As is evident from panels (b) and (c), the nucleation barrier and rate are not strongly dependent on the larger of a pore's dimensions, as long as that dimension is large enough. This can be understood by noting that nucleation within a pore occurs at a corner; as a droplet grows, it is stabilized energetically when it encounters the closer of the two other pore walls. Provided that the distance to the further wall is large enough, that distance does not strongly affect the ability of the nucleus to grow out into solution.

This observation suggests that it is preferable to pattern a surface by repeating in it copies of a well-chosen pore (a small pore that effects a substantial reduction in nucleation free energy barrier, such as the $10 \times 3$ pore), rather than e.g. etching long grooves in it. Consider, as one possible choice, the best square pore ($L_x=L_y=5$; see \fig{Fig:hot_spot_cuts}(a)). The larger of the in- and out barriers to nucleation for that pore is $\approx 20 \, \kt$. For a substrate row of (large) length $L$, we therefore expect the nucleation timescale associated with repeated, closely-spaced copies of the square to be $\tau_{\rm square} \sim (6/L) \exp(20)$.  By comparison, the barrier to nucleation for a periodic groove of the same width, built by imposing periodic boundaries in the $y$-direction of a box of length 30, is $24.7 \, \kt$, and so we expect the nucleation timescale for a groove of length $L$ to be $\tau_{\rm groove} \sim (30/L) \exp(24.7)$.  We therefore expect that replacing a single long groove of width 5 by an array of square pores of the same width will increase nucleation rate by a factor of $\tau_{\rm groove}/\tau_{\rm square} \sim 500$. Rate calculations done using forward flux sampling are consistent with this estimate. 

\subsection{Nucleation mechanisms in 3d pores also depend on pore size and aspect ratio}

\fig{Fig:hot_spot_breakdown} illustrates the range of behaviors associated with different pore shapes and sizes. For some pores, nucleation happens in a single step: 1) droplets are not stable within very small pores; instead, a critical nucleus appears on the surrounding surface (top left); or 2) droplets can attain criticality within very large pores (top right). For other pores, nucleation happens in two steps: 3) small filled pores are metastable with respect to both solution and the nucleated phase (bottom left); and 4) filled, moderately-sized pores are stable with respect to solution, and metastable with respect to the nucleated phase (bottom right).

\section{Conclusions}

We have studied homogeneous and heterogeneous nucleation in the 2d and 3d Ising models. Our key result is the extension of the work of Ref.~\cite{page2006heterogeneous} to calculate rates and free energy profiles for nucleation in the 3d Ising model in the presence of cuboidal pores. Pores of well-chosen aspect ratio can dramatically speed nucleation relative to a planar surface made of the same material, while badly-chosen pores provide no such enhancement. Further, for given thermodynamic conditions, and a sufficiently strong pore-particle attraction, there exists a pore size and aspect ratio ideal for promoting nucleation. \fig{Fig:behavior_range} summarizes the importance of pore choice in reducing free energy barriers to nucleation. A sufficiently attractive surface can dramatically reduce the nucleation barrier relative to that in bulk. For the parameters considered here, the bulk free energy nucleation barrier is about $67\, \kt$, while the barrier in the presence of an attractive wall ($J_{\rm s} = 0.6$) is about $45\, \kt$. A pore made from the same material can reduce the barrier even further: the barrier is about $20\, \kt$ for the optimally-sized square pore (\fig{Fig:behavior_range}(a)). However, a badly-chosen pore offers no improvement over a planar surface: although small droplets appear first within the pore of \fig{Fig:behavior_range}(b), the critical nucleus forms instead on the surface surrounding it.

\section{Acknowledgements}
We thank Richard Sear, Seunghwa Ryu and Wei Cai for valuable correspondence. L.O.H. was supported by the Center for Nanoscale Control of Geologic CO$_2$, a U.S. D.O.E. Energy Frontier Research Center, under Contract No. DE-AC02--05CH11231. This work was done at the Molecular Foundry, Lawrence Berkeley National Laboratory, supported under Contract No. DE-AC02-05CH11231. This research used resources of the National Energy Research Scientific Computing Center, which is supported by the Office of Science of the U.S. Department of Energy under Contract No. DE-AC02-05CH11231.

\appendix

\begin{figure*}[!t]
\center
\includegraphics[width=0.8 \linewidth]{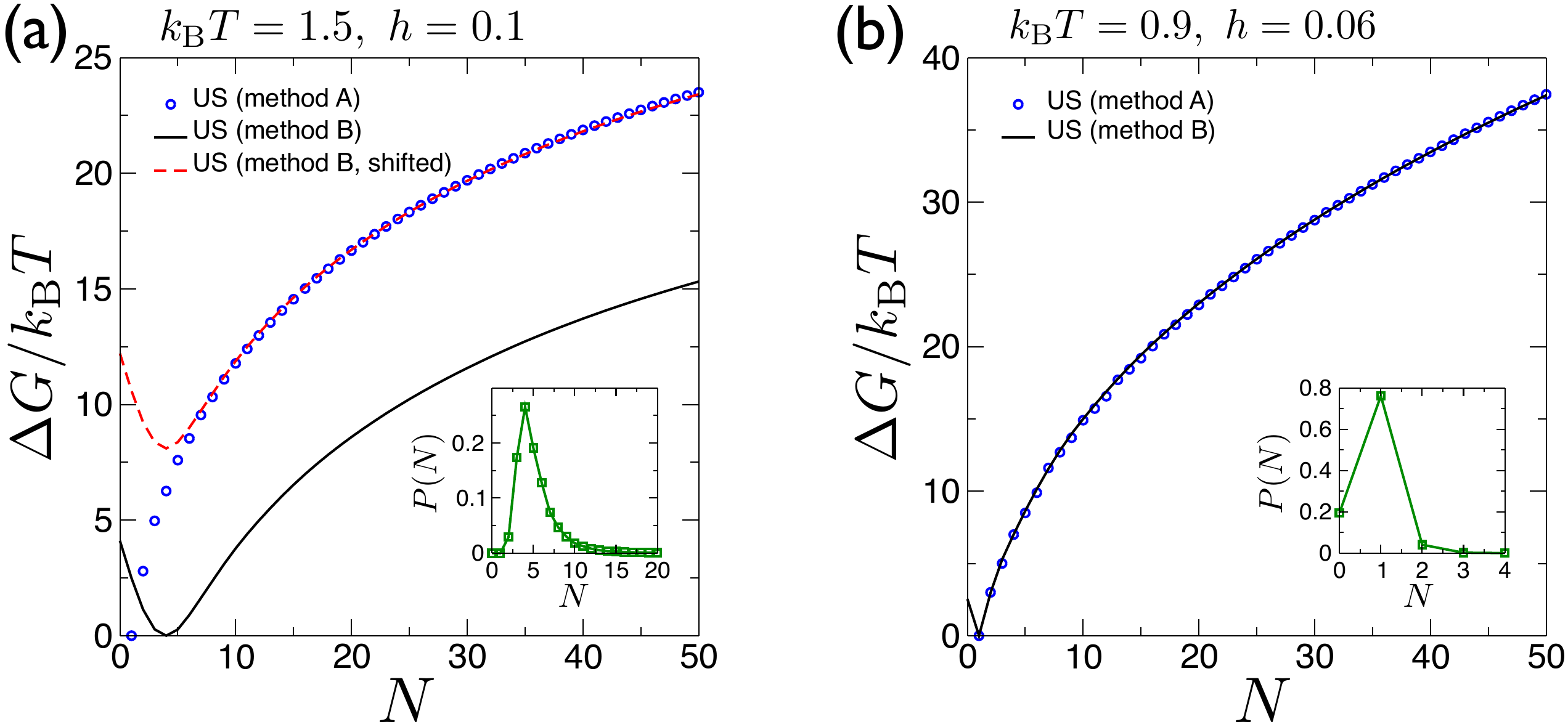}
\caption{Artifacts of measuring only the size of the largest cluster in the system are pronounced at deep supercooling but not at shallow supercooling. In panel (a) we show free energy profiles calculated by umbrella sampling methods A (measuring sizes of all clusters) and B (using as a reaction coordinate the system's largest cluster), under conditions that give rise to the smallest barrier in Fig~\ref{Fig:ising_bulk}(a). Here artifacts are apparent in the small-$N$ data produced by method B, because it ignores the many small clusters present in the simulation box (the inset, generated using unconstrained simulations, shows the likelihood that the largest cluster in the system is of size $N$). For larger $N$ the shape of the curves generated by the two methods agree, and when the `B' curve is shifted up by about $\ln V$~\cite{maibaum2008comment}, it sits on top of the `A' curve. (b) For conditions that give rise to the largest barrier in Fig.~\ref{Fig:ising_bulk}(b), by contrast, the curves generated by the two methods are similar, because the average cluster size is 1.}
\label{Fig:artifacts}
\end{figure*}

\section{Umbrella sampling}

\subsection{Sampling}
\label{apx:umbrella_sampling}

Several implementations of umbrella sampling~\cite{torrie1977nonphysical} for the study of nucleation are described in the literature~\cite{Pan:2004,wolde1999homogeneous, Bhimalapuram:2007}. We used a hybrid of the following two methods. In the first method (Method A) we measured the distribution of sizes of all connected clusters in the simulation box~\cite{Pan:2004,Maibaum:2008,maibaum2008comment}. We carried out `hard wall' umbrella sampling simulations~\cite{Pan:2004}, constraining the simulation to a `window' (of length 10 and with an overlap of 5 with its next neighbor) by rejecting any spin flip that made the largest cluster in the simulation box larger or smaller than the window's limits. Within each window we recorded histograms of the density of clusters of all sizes that fall within the window's bounds, measuring $\rho(N) = \langle M_N \rangle/V$, where $M_N$ is the number of clusters of size $N$ in the simulation box, and $V$ is its volume. We combined data from all windows using the weighted histogram analysis method (WHAM)~\cite{Ferrenberg:1989},  giving the free energy $G_{\rm A}(N) = -\kt \ln \rho(N)$. We checked this sampling procedure by reproducing, for the 3d Ising model, the curves shown in Fig. 5 of Ref.~\cite{Pan:2004} and Fig. 1 of Ref.~\cite{Maibaum:2008}.

While simple to implement, the hard wall umbrella sampling technique can become inefficient when there exist steep gradients in free energy. In this situation the sampling within a window is largely confined to the region adjacent to the wall lower in free energy, leaving the other end of the window badly sampled. Although it is possible to circumvent this problem by increasing the number of sampling windows (or their degree of overlap), this increases the computational cost of the method. A common solution (and the second method considered, Method B) is   to constrain the size $\nm$ of the largest cluster using a harmonic bias potential, $k_i(\nm-N^i_{\rm target})^2/2$~\cite{wolde1999homogeneous}. Here $i=1,2,...$ designate the different sampling windows, $k_i$ is a spring constant, and $N_{\rm{target}}^i$ is the cluster size at the center of window $i$. The width of a window is determined by the strength of the spring constant $k_i$ which can be softened or stiffened according to the local free energy gradient in order to improve sampling. We generated histograms $P_{\rm{B}}(\nm) = {\cal N}(\nm)/\sum_{\nm} {\cal N}(\nm)$ by adding 1 to a register ${\cal N}(\nm)$ if, after every trial move, the largest cluster in the system was of size $\nm$ (regardless of how many clusters of that size there were). To generate overlapping histograms we used a window spacing of $N^{i+1}_{\rm target}-N^i_{\rm target}=5$ and a spring constant of $k_i=0.2$. Sampling was done for a minimum of $10^6$ MC sweeps within each window. An MC sweep consisted of $N_{\rm{bulk}}$ attempted trial moves, where $N_{\rm{bulk}}$ was the number of lattice sites in the bulk of the simulation box, i.e. sites that are not part of a wall or pore. We updated and recorded cluster size distributions after every trial move, so maximizing the efficiency of our simulations. We used umbrella integration~\cite{Kastner:2005} to unbias the results of umbrella sampling simulations, and used WHAM~\cite{Ferrenberg:1989} to check this procedure and resolve fine details of certain free energy curves (see Appendix~\ref{apx:umbrella_integration}). This procedure gives the `free energy' $G_{\rm B}(\nm) = -\kt \ln P_{\rm B}(\nm)$. 

Unlike method A, which measures the distribution of sizes for all clusters, method B instead samples the probability that the \emph{largest} cluster is of size $\nm$.  As pointed out by Maibaum~\cite{maibaum2008comment}, a free energy penalty is incurred whenever $\nm$ is constrained to sizes smaller than the average cluster size, $N_{\rm{av}}$, seen in unconstrained simulations (that do not result in nucleation). \fig{Fig:artifacts} shows a comparison between the two umbrella sampling methods at conditions of deep (a) and shallow (b) supercooling. Here $\Delta G$ is defined as $G_{\rm{A}}(N)-G_{\rm{A}}(1)$ and $G_{\rm{B}}(\nm) - \min(G_{\rm{B}}(\nm))$ for methods A and B respectively. At deep supercooling the average cluster size at small $N$ is 5; consequently, a spurious increase in $\Delta G$ is seen in the free energy curve obtained using method B for $N<5$~\cite{maibaum2008comment}. In contrast, at shallow supercooling the average cluster size at small $N$ is 1, and the methods agree.

Although small-$N$ artifacts can be present using method B, it is important to note that the shape of the free energy curve is correct for $N \gg N_{\rm av}$. We can therefore take advantage of the improved sampling that method B provides by using it to sample large clusters, and using method A to gather statistics for small clusters (which is typically cheap to do). The curves reported in the text were stitched together using both methods, with the method B results shifted vertically to match the small-cluster data obtained using method A. Comparison with the CNT-like predictions in Section~\ref{Section:ising_bulk} and the results of Ref.~\cite{Ryu:2010} allow us to confirm that this scheme works. 

To complement our free energy sampling we calculated nucleation rates directly using the forward flux sampling (FFS) method~\cite{Allen:2006}. FFS is reasonably insensitive to the choice of reaction coordinate~\cite{Allen:2006}, and we verified that consistent results were obtained using as a reaction coordinate 1) the size of the largest cluster in our simulation box and 2) the total number of particles. All interfaces were spaced 10 particles apart, and 10000 crossings were stored at each. Rates were measured in units of Monte Carlo steps per bulk site. 

To measure nucleation rates in the presence of pores we followed the protocol outlined in Ref.~\cite{page2006heterogeneous} and decomposed the overall nucleation rate into two parts: the rate for pore filling $R_{\rm{in}}$; and the rate for nucleation out of an already filled pore $R_{\rm{out}}$. These rates define the mean timescales for the two nucleation processes, $\tau_{\rm{in}}=R_{\rm{in}}^{-1}$ and  $\tau_{\rm{out}}=R_{\rm{out}}^{-1}$, which can be combined to give the overall nucleation time, $\tau_{\rm{total}}=\tau_{\rm{in}} + \tau_{\rm{out}}$, and hence the overall nucleation rate $R_{\rm{total}} = \tau_{\rm{total}}^{-1}$.

\subsection{Unbiasing distributions}
\label{apx:umbrella_integration}
We combined data from each window using the weighted histogram analysis method (WHAM)~\cite{Ferrenberg:1989} and/or umbrella integration~\cite{Kastner:2005}. WHAM takes as its input an overlapping sequence of probability distributions, and effects a self-consistent iteration to reconstitute the underlying free energy curve. Umbrella integration, by contrast, assumes Gaussian probability distributions within each window, and takes as input only the mean and variance of the sampled distribution. It also does not require windows to overlap. It is therefore computationally cheaper to implement than WHAM. For example, the contour plots shown in~\fig{Fig:hot_spot} were made from 210 sets of simulations, each comprising 200 individual sampling windows. Using umbrella integration it was possible to compute all of the 210 free energy curves in less than 10 seconds. By contrast, WHAM took north of 20 minutes to process an individual free energy curve.
\begin{figure}[!b]
\center
\includegraphics[width=0.9\columnwidth]{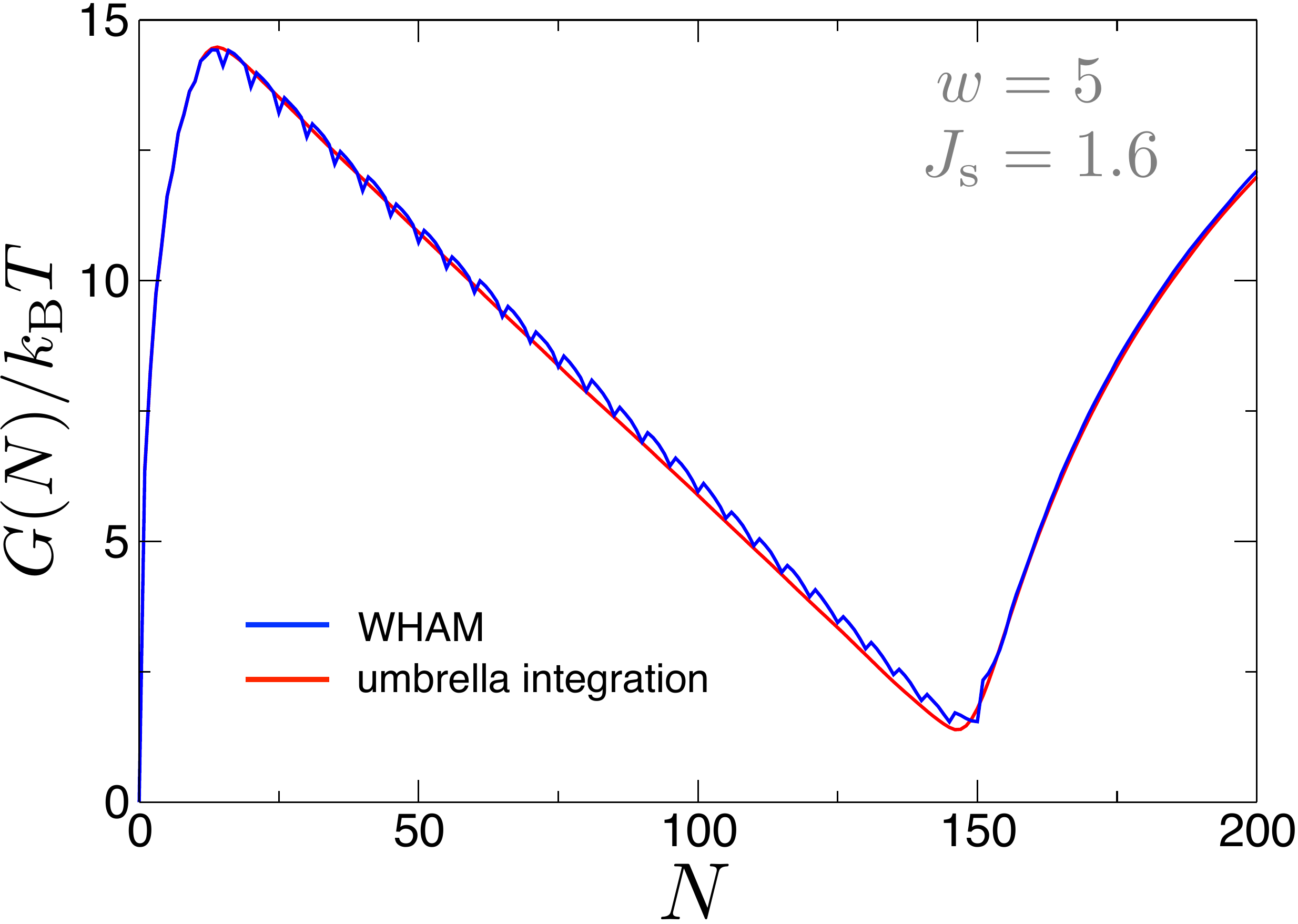}
\caption{An enlargement of the boxed region shown in~\fig{Fig:ising_pore}(a). We used umbrella integration as a fast way to reconstitute the large-scale features of free energy profiles, and WHAM to resolve profiles' small features when probability distributions within windows were non-Gaussian.}
\label{Fig:wham_vs_ui}
\end{figure}

As the authors of the method caution~\cite{Kastner:2005}, umbrella integration misses features of a free energy curve if the sampling within windows is non-Gaussian, as happens when a pore fills layer by layer. In such cases, WHAM reveals subtle local features (metastable minima) in the free energy profile: see e.g. the $w=5$ profile in \fig{Fig:ising_pore}(a), enlarged in \fig{Fig:wham_vs_ui} for clarity. Umbrella integration does not. However, umbrella integration gets correct the overall shape of the free energy profile, including the positions and heights of the large free energy barriers (corresponding to the initial nucleation event within the pore, and nucleation of a droplet out of the filled pore). We therefore used umbrella integration to broadly survey free energy landscapes in Figs~\ref{Fig:hot_spot}--\ref{Fig:hot_spot_breakdown}, and used WHAM to resolve subtle features of particular curves.

\section{Nucleation barriers at a planar surface are reasonably well described by CNT}
\label{sec:plane}
\begin{figure*}[!tb]
\center
\includegraphics[width=0.8\linewidth]{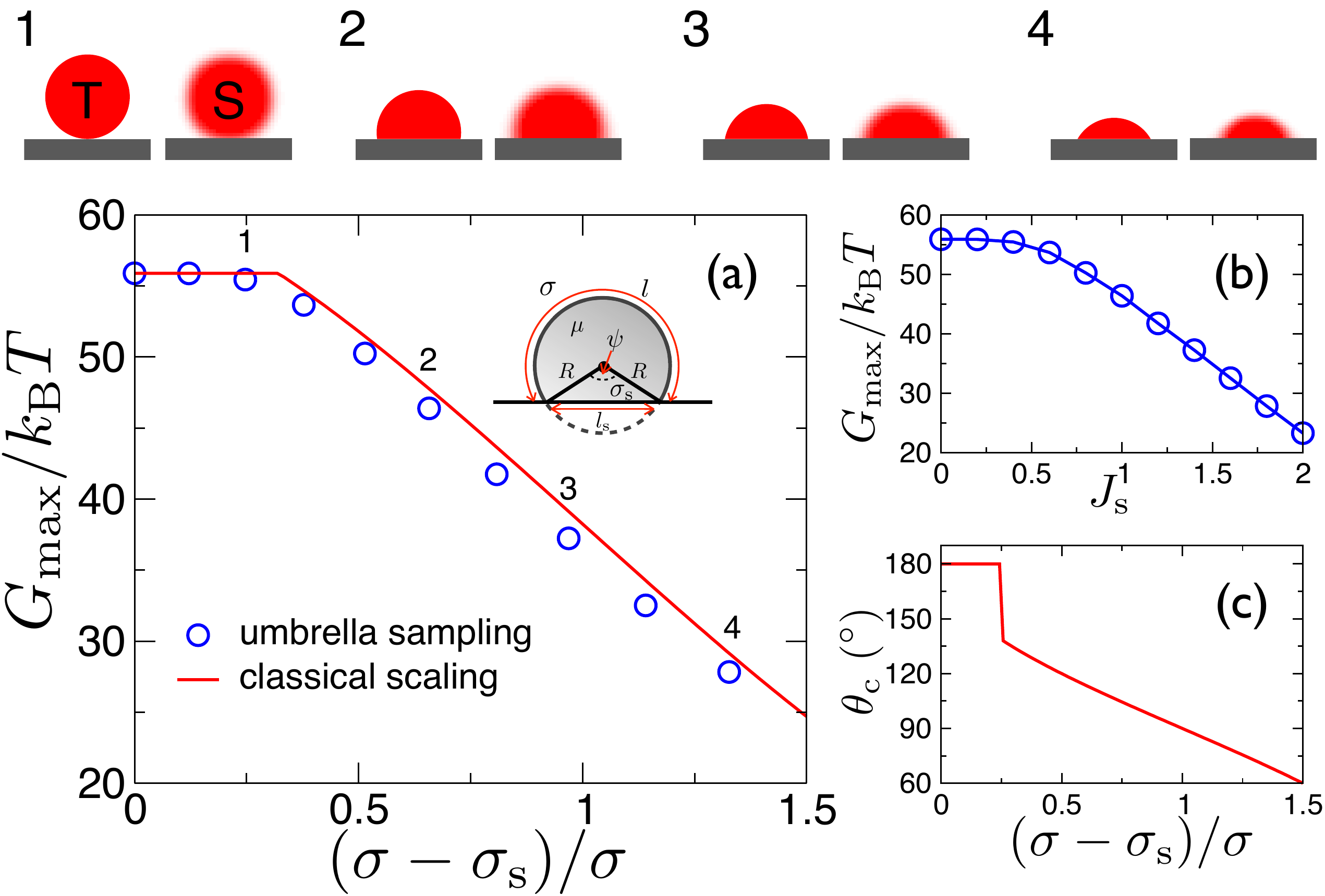}
\caption{(a) Free energy barriers to nucleation in the 2d Ising model are well-described by CNT. We show free energy barriers as a function of reduced surface tension computed from umbrella sampling simulations (black) and the classical scaling approximation (red) of \eqq{Eqn:free_energy_wall_shifted}. The bulk free energy barrier (55.9 $\kt$) is used whenever \eqq{Eqn:free_energy_wall_shifted} returns a value larger than it. There is a reasonable agreement between simulation and theory. (b) Free energy barrier from umbrella sampling simulations as a function of the particle-wall interaction strength $J_{\rm{s}}$. As shown in \fig{Fig:ising_surface}, nucleation occurs at the wall only if the particle-wall attraction is strong enough to overcome the entropic cost of moving the droplet to the wall. (c) Contact angle vs reduced surface tension computed from Young's equation. The contact angle is $180^{\circ}$ whenever bulk nucleation is preferred. The four numbered snapshots show the average profiles of critical nuclei at different values of the reduced surface tension. Those on the left are computed from Young's equation, while those on the right are averages from umbrella sampling simulations. The qualitative agreement between theory (T) and simulation (S) is good.}
\label{Fig:ising_surface2}
\end{figure*}

The requirement that a wall attract a nucleating phase before bulk nucleation is suppressed is consistent with a classical nucleation theory-like scaling argument. We can estimate the barrier for nucleation at a wall by considering a droplet whose shape is the portion of a circle of radius $R$ that lies above a surface when the surface-circle intersections subtend an angle $\psi$ at the center of the circle (see \fig{Fig:ising_surface2}, inset). When $\psi=0$ the circle just touches the wall and we have nucleation in the bulk; when $\psi=\pi$ the droplet is a semi-circle; when $\psi=2 \pi$ we consider the droplet to wet the wall. We estimate the free energy of the droplet as 
\begin{equation}
\label{Eqn:cnt_wall}
	G_{\rm{s}}(R,\psi)=-A \Delta g + \sigma l+  \sigma_{\rm s} l_{\rm s}+ \frac{1}{2} \Theta(\psi) \ln N_{\rm box}.
\end{equation}
The first three terms account for  droplet area $A=N=(\pi-\psi/2) R^2+(1/2) R^2 \sin \psi$; curved droplet perimeter $l = (2 \pi -\psi) R$; and wall-contacting droplet perimeter $l_{\rm s} = 2 R \sin(\psi/2)$. $\sigma_{\rm s}$ is the wall-droplet surface tension. The final term ($\Theta(\psi) =1$ if $\psi >0$, and is zero if $\psi=0$) accounts for the fact that there are more ways of placing a droplet in the bulk than at the wall of the system, i.e. there is an entropic cost associated with moving a cluster from the bulk to the wall. Maximizing \eqq{Eqn:cnt_wall} with respect to $R$ gives the critical radius 
\begin{equation}
\label{Eqn:critical_radius_wall}
	R_{\rm{c}}(\psi) = \frac{\sigma\left(2\pi - \psi\right) + 2\sigma_{\rm{s}} \, \mathrm{sin}(\psi/2)}{\Delta g\left(2\pi - \psi + \mathrm{sin}(\psi)\right)}.
\end{equation}
Substituting this expression into \eqq{Eqn:cnt_wall} gives the free energy barrier to nucleation:
\bea
\label{Eqn:free_energy_wall}
	G_{\rm s}(R_{\rm{c}},\psi) &=& \frac{\left[\sigma\left(2\pi - \psi\right) + 2\sigma_{\rm{s}} \mathrm{sin}(\psi/2)\right]^2}{2 \Delta g \left(2\pi - \psi + \mathrm{sin}(\psi)\right)} \nonumber \\
	&+&\frac{1}{2} \Theta(\psi) \ln N_{\rm box}.
\eea
Upon setting $\psi=0$, we recover the conventional bulk solution,~\eqq{Eqn:ising_free_energy_corrected} with $\tau=d=0$. The droplet makes a contact angle with the surface of $\theta = \pi-\psi/2$. Balancing surface tensions using Young's equation~\cite{Young:1805} gives $\sigma \, \mathrm{cos}\theta=-\sigma_{\rm{s}}$. Using these results in \eqq{Eqn:free_energy_wall} gives a contact angle-dependent free energy barrier
\begin{equation}
\label{Eqn:free_energy_theta2}
	G^{\rm s}_{\rm max} = f(\theta) G_{\rm{bulk}}(R_{\rm{c}}^{\rm{bulk}})+ \frac{1}{2}\Theta(\pi - \theta) \ln N_{\rm box},
\end{equation}
where
\begin{equation}
\label{Eqn:f_theta}
	f(\theta) =  \frac{1}{\pi}\left(\theta - \frac{1}{2}\mathrm{sin}(2\theta)\right).
\end{equation}
This is the Turnbull estimate for the free energy of a 2d droplet at a surface~\cite{Turnbull:1950}. In terms of surface tensions the function $f$ reads
\begin{equation}
\label{Eqn:f_sigma}
	f(\sigma,\sigma_{\rm{s}}) = \frac{1}{\pi}\left[\frac{\sigma_{\rm{s}}}{\sigma^2}\sqrt{(\sigma - \sigma_{\rm{s}})(\sigma+\sigma_{\rm{s}})} + \mathrm{arccos}(-\sigma_{\rm{s}}/\sigma)\right].
\end{equation}
To compare this estimate with the results of simulation we added to \eqq{Eqn:free_energy_theta2} the difference between the numerical value for the bulk free energy barrier, calculated via umbrella sampling, and the uncorrected CNT prediction, i.e.
\begin{equation}
\label{Eqn:free_energy_wall_shifted}
	G^{\rm s}_{\rm max} \to f(\theta) G_{\rm{bulk}}(R_{\rm{c}}^{\rm{bulk}})+ \frac{1}{2} \Theta(\pi - \theta) \ln N_{\rm box} + \Delta G_{\rm{bulk}},
\end{equation}
where  $\Delta G_{\rm{bulk}} = G_{\rm{bulk}}^{\rm{sim}} - G_{\rm{bulk}}^{\rm{CNT}}|_{d=\tau=0}$. This modification assumes that the ill-defined origin of our CNT-like expression can be fixed by requiring that in the bulk limit it returns the results of computer simulations (see Section \ref{Section:ising_bulk}). This modification is ad-hoc and uncontrolled.

We take Onsager's solution for the Ising model surface tension, \eqq{Eqn:onsager_sigma}, as an approximation~\footnote{For the bulk conditions used here the Onsager approximation lies within 2\% of the Shneidman effective surface tension, and we choose to use the former for simplicity. Under other conditions it may be desirable to replace this equation with the expression for $\sigma_{\rm{eff}}(T)$ given in~\cite{Shneidman:1999}.} for the droplet-solution surface tension:
\begin{equation}
\label{Eqn:sigma_lattice_gas}
	\sigma = J/2 - \kt \ln \left[\mathrm{coth}(\beta J/4)\right].
\end{equation}
Note that $J$ is the lattice gas coupling, not the Ising one. To estimate a value for the droplet-wall surface tension, we note that the two terms in the Onsager surface tension account for the energy per unit length of a planar interface (first term), and the free energy per unit length of fluctuations normal to that interface (second term)~\cite{Kardar:2007}. To see this, consider an interface of horizontal length $L$ between up- and down spins in a 2d Ising model (at $h=0$), where the vertical position of the interface at the $n^{\rm th}$ lattice site across it ($n=1,2,\dots,L)$ is $u_n$. The energy cost of fluctuations normal to the interface is therefore $2 K \sum_{n=1}^L |u_{n+1}-u_n|$, and the associated partition function is
\bea
Z_{\rm interface} &=& \sum_{\{u_n\}} {\rm e}^{-2 \beta K \sum_n |u_{n+1}-u_n|}  \nonumber \\
&=& \left( \sum_{\Delta u=-\infty}^\infty {\rm e}^{-2 K \beta | \Delta u |}\right)^L  \nonumber \\
&=& \coth^L \left( \beta K\right).
\eea
The free energy per unit length associated with perpendicular fluctuations of the interface, $f_{\rm interface}=-L^{-1} \kt \ln Z_{\rm interface}$, is therefore the second term in \eqq{Eqn:sigma_lattice_gas} (recall that $K=J/4$). We now guess that when the wall is attractive enough that a droplet remains in close contact with it, the wall-droplet interface acts {\em as if} the wall can fluctuate. Clearly this is not true microscopically, and is likely to be a poor approximation if the droplet surface moves appreciable away from the wall. We nonetheless conjecture that the surface tension between the droplet and the wall can be approximated as
\begin{equation}
\label{Eqn:sigma_surface}
	\sigma_{\rm{s}} = (J-J_{\rm{s}})/2 - \kt \ln \coth(\beta(J-J_{\rm{s}})/4).
\end{equation}
We shall use Eqns.~\eq{Eqn:sigma_lattice_gas} and \eq{Eqn:sigma_surface} to relate the surface tensions entering \eqq{Eqn:f_sigma} to the particle-wall attraction $J_{\rm s}$ used in our simulations. 
Finally, our CNT-like prediction for the barrier to nucleation in the presence of a surface is 
\beq
G_{\rm max}=\min \left(G^{\rm s}_{\rm max},  G_{\rm bulk}^{\rm{sim}} \right),
\eeq
i.e. we take the bulk barrier whenever the barrier to nucleation at the surface is larger than it (by virtue of the entropic penalty of wall confinement). \fig{Fig:ising_surface2}(a) shows a comparison between  theoretical (solid line) and simulated (circles) free energy barriers. Both calculations show that  nucleation only occurs at the wall when the particle-wall attraction is strong enough to offset the entropic penalty of removing the droplet from the bulk. Moreover, the quantitative agreement between the two methods is reasonable, which is surprising in light of the crude nature of the approximations we have made. \fig{Fig:ising_surface2}(b) shows the umbrella sampling free energy barrier as a function of the particle-wall interaction strength. \fig{Fig:ising_surface2}(c) shows the contact angle computed from Young's equation, which is $180^{\circ}$ whenever bulk nucleation is preferred. The numbered snapshots show the average profiles of critical nuclei at different values of the reduced surface tension. Those on the left are generated using the contact angle from the solution to Young's equation, while those on the right are averages from umbrella sampling simulations. In all cases the agreement between theory and simulation is good.  Here we have made a pictorial comparison between droplet shapes predicted by theory and computed by simulation; previous work demonstrates quantitatively that CNT can predict the Ising model droplet contact angle~\cite{winter2009monte,winter2009heterogeneous}.

\end{document}